\RequirePackage{amsmath}
\documentclass[runningheads,a4paper]{llncs}

\input glyphtounicode.tex %

\usepackage[utf8]{inputenc}
\usepackage{amsfonts}
\usepackage{graphicx}
\usepackage{subfigure}
\usepackage{float}

\usepackage[retainorgcmds]{IEEEtrantools}
\usepackage[usenames,dvipsnames]{xcolor}
\usepackage{color}

\usepackage{bm}
\usepackage{hyperref}
\usepackage{graphicx}
\usepackage{amssymb}
\usepackage{lineno}
\usepackage{multirow}
\usepackage{setspace} 
\usepackage{booktabs}
\usepackage{amsmath}
\usepackage{float}
\usepackage{color}
\usepackage{multirow}

\begin{document}

\mainmatter  

\title{3D Cardiac Shape Prediction with Deep Neural Networks: Simultaneous Use of Images and Patient Metadata}
\titlerunning{3D Cardiac Shape Prediction with Deep Neural Networks}

\authorrunning{R.~Attar et al.}
\author{Rahman~Attar$^{1,2}$, Marco~Perea\~nez$^{1}$, Christopher Bowles$^{1}$, Stefan~K.~Piechnik$^3$, Stefan~Neubauer$^3$, Steffen~E.~Petersen$^{4}$, Alejandro~F.~Frangi$^{1,2}$}
\institute{$^1$Center for Computational Imaging and Simulation Technologies in Biomedicine, School of Computing, University of Leeds, Leeds, UK.\\$^2$ Biomedical Imaging Department, Leeds Institute for Cardiovascular and Metabolic Medicine, School of Medicine, University of Leeds, Leeds, UK. \\\email{\{r.attar, a.frangi\}@leeds.ac.uk}\\$^3$Oxford Center for Clinical Magnetic Resonance Research (OCMR), Division of Cardiovascular Medicine, University of Oxford, John Radcliffe Hospital, Oxford, UK.\\ $^4$William Harvey Research Institute, NIHR Barts Biomedical Research Unit, Queen Mary University of London and Barts Heart Centre, St Bartholomew's Hospital, Barts Health NHS Trust, London, UK}

\maketitle
\begin{abstract}
Large prospective epidemiological studies acquire cardiovascular magnetic resonance (CMR) images for pre-symptomatic populations and follow these over time. To support this approach, fully automatic large-scale 3D analysis is essential. In this work, we propose a novel deep neural network using both CMR images and patient metadata to directly predict cardiac shape parameters. The proposed method uses the promising ability of statistical shape models to simplify shape complexity and variability together with the advantages of convolutional neural networks for the extraction of solid visual features. To the best of our knowledge, this is the first work that uses such an approach for 3D cardiac shape prediction. We validated our proposed CMR analytics method against a reference cohort containing 500 3D shapes of the cardiac ventricles. Our results show broadly significant agreement with the reference shapes in terms of the estimated volume of the cardiac ventricles, myocardial mass, 3D Dice, and mean and Hausdorff distance. 
\end{abstract}

\section{Introduction}
Cardiovascular disease (CVD) is the most prevalent cause of death worldwide~\cite{attar}. Early quantitative assessment of cardiac function and structure allow for proper preventive care, and early cardiovascular treatment. To support such an approach, analysis and interpretation of large-scale population-based cardiovascular magnetic resonance (CMR) imaging studies are of high importance in the medical image analysis community. This helps to identify  patterns and trends across population groups, and accordingly, reveal insights into key risk factors before CVDs fully develop.

We believe that true 3D analysis is essential for the structural assessment of global and regional cardiac function. We propose a new approach that ensures the global coherence of the cardiac anatomy and naturally lends itself to further analysis in which full 3D anatomy is necessary; for example, in mechanical and flow simulations, or modelling the relationship between cardiac morphology and patient information such as: socio-demographic, lifestyle and environmental, family  history, genetic, and omics data. 

Though fully automatic 3D segmentation is required for further analysis, the complexity of anatomical structures and their local intensity variation across a population cohort make it challenging. Statistical 3D shape model-based approaches such as \cite{attar} have been successfully used for automatically segmenting cardiac structures and generating associated function indexes. This is mainly attributed to the inclusion of prior knowledge of the cardiac shape into the segmentation method. These segmentation approaches typically use very simple features such as gradients on intensity profiles to fit a 3D model. This is an iterative process in which the goal is to minimise the Mahalanobis distance between an intensity profile sampled at a candidate position and its corresponding intensity appearance model by deforming the shape within its range of normal variation to match the image data. On the other hand, in the last decade, fully convolutional networks (FCN) have shown great potential in image-based pattern recognition in a variety of tasks, including cardiac segmentation. However, their output results are, by nature, 2D segmentations masks for every short axis (SAX) and long axis (LAX) CMR slices. Although these 2D masks are sometimes extended via a further step of non-rigid registration to a 3D atlas to produce a 3D cardiac shape~\cite{duan2019automatic}, this is not efficient for learning topological shape information.

In this paper, we propose to exploit image features obtained using deep FCNs trained on both SAX and LAX views, along with the rich shape priors learned using statistical shape models, to jointly and simultaneously predict the parameters of 3D cardiac shapes, instead of a pixel-wise classification across each 2D slice. Another significant aspect of this work is the integration of patient metadata into the process of shape prediction using a Multilayer Perceptron (MLP). This information, which is currently ignored in cardiac segmentation or shape generation, has been shown in different clinical studies to have an impact on cardiac morphology and structure~\cite{gilbert2019independent}. We hope this work inspires other researchers to exploit the priors offered by patient metadata in other applications for potentially more accurate and patient-specific models. 

\textbf{The contributions of this paper are three-fold:} we propose 
 1) an innovative end-to-end deep neural network that directly predicts 3D shape parameters derived from a Principal Component Analysis (PCA) space;
 2) a novel approach using two CMR image views and patient metadata simultaneously to predict cardiac shape; 
 3) a creative loss function defined in the domain of 3D shape parameters which weights each PCA mode of variation independently, prioritising the more significant modes and leading to more accurate shape prediction.
 
\section{Methods}
\subsection{Reference 3D Shapes of Cardiac Ventricles} \label{Reference3DShapes}
\begin{figure*}[!ht]
	\centering\includegraphics[width=1\linewidth]{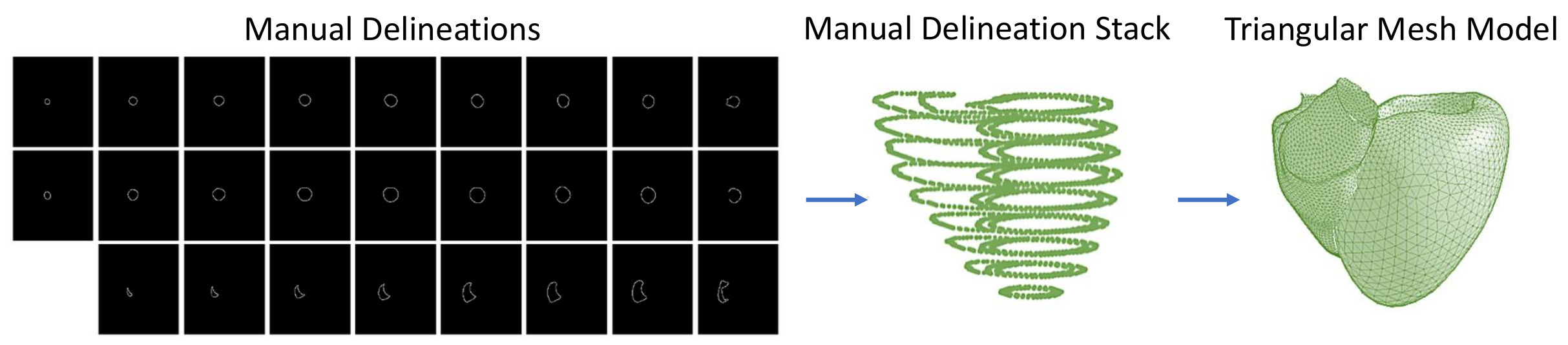}
	\caption{An example 3D shape of cardiac ventricles constructed from a stack of 2D manual contours on SAX view slices.}
		\label{registration}
\end{figure*}

We generated a reference cohort of 3D shapes through the non-rigid registration of a 3D biventricular model to a set of 3D points obtained from manual delineations using the Coherent Point Drift (CPD) method~\cite{myronenko2010point}. The 3D model is comprised of two structures; the Left Ventricle (LV) and the Right Ventricle (RV). The LV is a closed water-tight mesh comprising both endo and epicardial walls. The RV is an open mesh representing only the RV endocardium. The RV has two openings, the atrioventricular valve opening, and the pulmonary valve opening. Figure~\ref{registration} shows a sample of manual 2D contours and its corresponding 3D shape obtained from the CPD method.
\vspace{-0.5cm}

\vspace{-1cm}
\subsection{Point Distribution Mode (PDM)}
The PDM encodes the mean and variance of the 3D cardiac shapes. The PDM is constructed during training using PCA on a set of generalised Procrustes-aligned shapes. Assume a training set of $M$ shapes, each described by $N$ points in $\mathcal{R}^3$, i.e., $(\mathbf{x}_j^i, \mathbf{y}_j^i, \mathbf{z}_j^i)$ with $i=1, ..., M$ and $j=1, ..., N$. Further, let $\mathbf{s}_i = (\mathbf{x}_1^i, \mathbf{y}_1^i, \mathbf{z}_1^i, ..., \mathbf{x}_N^i, \mathbf{y}_N^i, \mathbf{z}_N^i)^T$ be the $i$-th vector representing the $i$-th shape. Finally, let $\mathbf{S} = [\mathbf{s}^1, ..., \mathbf{s}^M]$ be the set of all training shapes in matrix form. The shape class mean and covariance of $\mathbf{S}$ is calculated as follows:
\vspace{-0.15cm}
\begin{equation}
\bar{\mathbf{s}} = \dfrac{1}{M}\sum_{i=1}^{M}\mathbf{s}_i ~~~~~\mathrm{and}~~~~~ \mathbf{C} = \dfrac{1}{M-1}\sum_{i=1}^{M}(\mathbf{s}_i-\bar{\mathbf{s}})(\mathbf{s}_i-\bar{\mathbf{s}})^T
\label{mean}
\vspace{-0.15cm}
\end{equation}

The shape covariance is represented in a low-dimensional PCA space. This provides $l$ eigenvectors $\mathbf{\Phi}= [\mathbf{\varphi}_1\mathbf{\varphi}_2...\mathbf{\varphi}_l]$, and corresponding eigenvalues $ \mathbf{\Lambda} = diag(\mathbf{\lambda}_1, \mathbf{\lambda}_2,...,\mathbf{\lambda}_l)$ computed through the Singular Value Decomposition of the covariance matrix . Hence, assuming the shape class follows a multi-dimensional Gaussian probability distribution, any shape in the shape class can be approximated from the following linear generative model:
\vspace{-0.15cm}
\begin{equation}
\mathbf{s} \approx \bar{\mathbf{s}} + \mathbf{\Phi}\mathbf{b}
\label{finalShape}
\vspace{-0.15cm}
\end{equation}
\noindent where $\mathbf{b}$ are shape parameters restricted to $|\mathbf{b}_i| \leq \beta \sqrt{\lambda_i}$; we typically set $\beta = 3$ to capture $99.7\%$ of shape variability. The shape parameters of $\mathbf{s}$ can then be estimated as follows:
\begin{equation}
\mathbf{b} = \mathbf{\Phi}_l^T (\mathbf{s}-\bar{\mathbf{s}}).
\label{b_fomula}
\end{equation}
\noindent Here, the entries of \textbf{b} are the projection coefficients of mean-centred shapes $(\mathbf{s} - \bar{\mathbf{s}})$ along the columns of $\mathbf{\Phi}$. 

\subsection{Images and Metadata}
Each CMR image volume was pre-processed as follows. Each 9-slice SAX stack was intensity normalised by saturating the top 0.2\% of intensities and scaling between 0 and 1, and spatially normalised by aligning a fixed point, defined as the average point of intersection between the three LAX views and each SAX slice, and angle, defined as the angle of the 4-chamber LAX, to a standard location and angle respectively. A 64$\times$64~px region of interest (ROI) was then sampled from each slice at a 2~mm isotropic resolution. The corresponding 4-chamber LAX views were similarly intensity normalised, with a 80$\times$60~px  ROI sampled at the same 2~mm resolution around the point of intersection between the three LAX views and the basel SAX slice, and zero-padded to 80$\times$80~px. Table~\ref{metadata} shows the summary of the metadata available for every image volume including both continuous and categorical variables. All variables were scaled to the range [0, 1], including categorical variables (viz. \textit{sex}/\textit{alcohol} $\in (0,1)$, \textit{smoking} $\in (0,0.5,1)$).
\vspace{-0.5cm}

\begin{table*}[!h]
\caption{Summary of the subject metadata used in this study.}
	\centering
		\scalebox{0.6}{

\begin{tabular}{l|l|c}
\textbf{Type}                         & \textbf{\hspace{0.2cm}Metadata}               & \textbf{\hspace{0.2cm}Range}         \\ \hline
\multirow{8}{*}{\textbf{Continuous}}   & \hspace{0.2cm}Age (years)                     & \hspace{0.2cm}61~$\pm$~7                 \\
                                      & \hspace{0.2cm}Weight (kg)                     & \hspace{0.2cm}76~$\pm$~15                \\
                                      & \hspace{0.2cm}Height (cm)                     & \hspace{0.2cm}170~$\pm$~9                \\
                                      & \hspace{0.2cm}Body mass index (kg/m2)         & \hspace{0.2cm}27~$\pm$~4                 \\
                                      & \hspace{0.2cm}Body surface area (m2)          & \hspace{0.2cm}1.8~$\pm$~0.2              \\
                                      & \hspace{0.2cm}Heart rate (bpm)                & \hspace{0.2cm}68~$\pm$~11                  \\
                                      & \hspace{0.2cm}Diastolic blood pressure (mmHg) & \hspace{0.2cm}79~$\pm$~11                  \\
                                      &\hspace{0.2cm}Systolic blood pressure (mmHg)  & \hspace{0.2cm}139~$\pm$~19                 \\ \hline
\multirow{3}{*}{\textbf{Categorical}} & \hspace{0.2cm}Sex                             & \hspace{0.2cm}male$/$female            \\
                                      & \hspace{0.2cm}Smoking status                  & \hspace{0.2cm}never$/$previous$/$current \\
                                      & \hspace{0.2cm}Alcohol consumed                & \hspace{0.2cm}yes$/$no               
\end{tabular}}
\label{metadata} 
\vspace{-1cm}
\end{table*}

\subsection{Network Architecture and Loss Function}
\begin{figure*}[!ht]
	\centering\includegraphics[width=0.8\linewidth]{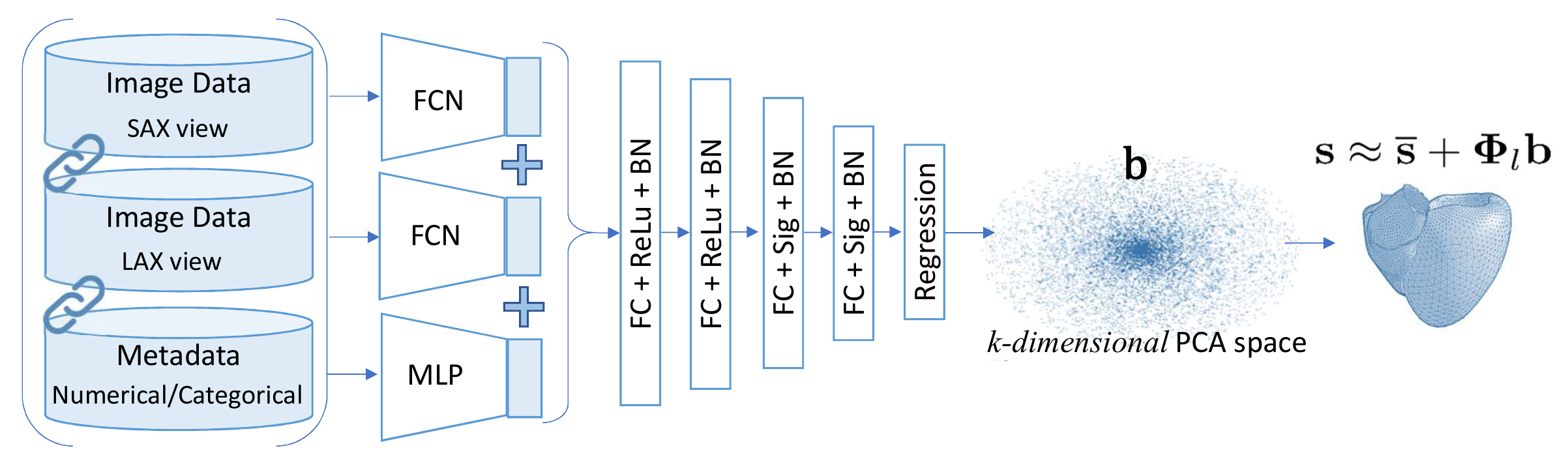} \vspace{-0.3cm}
	\caption{The proposed method extracts a high-level representation of the image from SAX and LAX views using two FCNs, and concatenates the image features together along with the output of an MLP network applied to the metadata. Four fully connected layers with ReLU or Sigmoid activation functions and batch normalisation then produce the $k$ parameters in PCA space which describe the 3D shape of the cardiac ventricles.}
		\label{CNNMLP}
		\vspace{-0.2cm}
\end{figure*}

Fig.~\ref{CNNMLP} shows a diagram of the proposed method. The network has three inputs: SAX view images, LAX view image, and metadata. The output is the predicted shape parameters $b^{P} = \{b^P_i |i=1, ..., k \}$. To train the proposed architecture, we introduce the following loss function for training: 
\vspace{-0.3cm}
\begin{equation}
\mathbf{E(\theta)} = \sum_{i=1}^{k} ~f(b^P_i (\theta),~b_i^{R})~.~w(i, k) ~~~~\mathrm{where} ~~~~ w(i, k)=\sqrt{\frac{k-i+1}{k}}
\label{loss}
\vspace{-0.15cm}
\end{equation} 
\noindent where $k$ is the number of shape parameters, $\theta$ denotes the network parameters, $f(.)$ denotes the absolute error of the difference between the reference value ($b_i^{R}$) and the value ($b^P_i (\theta) $) predicted by the network. $w(.)$ denotes a weighting function depending on the importance of the $i-th$ mode of variation on shape prediction, i.e. it assigns a higher weight to first modes of variation in the shape's PCA space. The first modes of variations in the PCA space are critical as they are the main parameters to affect the shape structure. Predicting these accurately is, therefore, more important as they have the greatest control over the final predicted shape. Ultimately, having the mean shape, eigenvectors and predicted shape parameters, the final shape can be predicted using Eq. \ref{finalShape}.
\vspace{-0.5cm}

\begin{figure*}[!ht]
	\centering\includegraphics[width=0.8\linewidth]{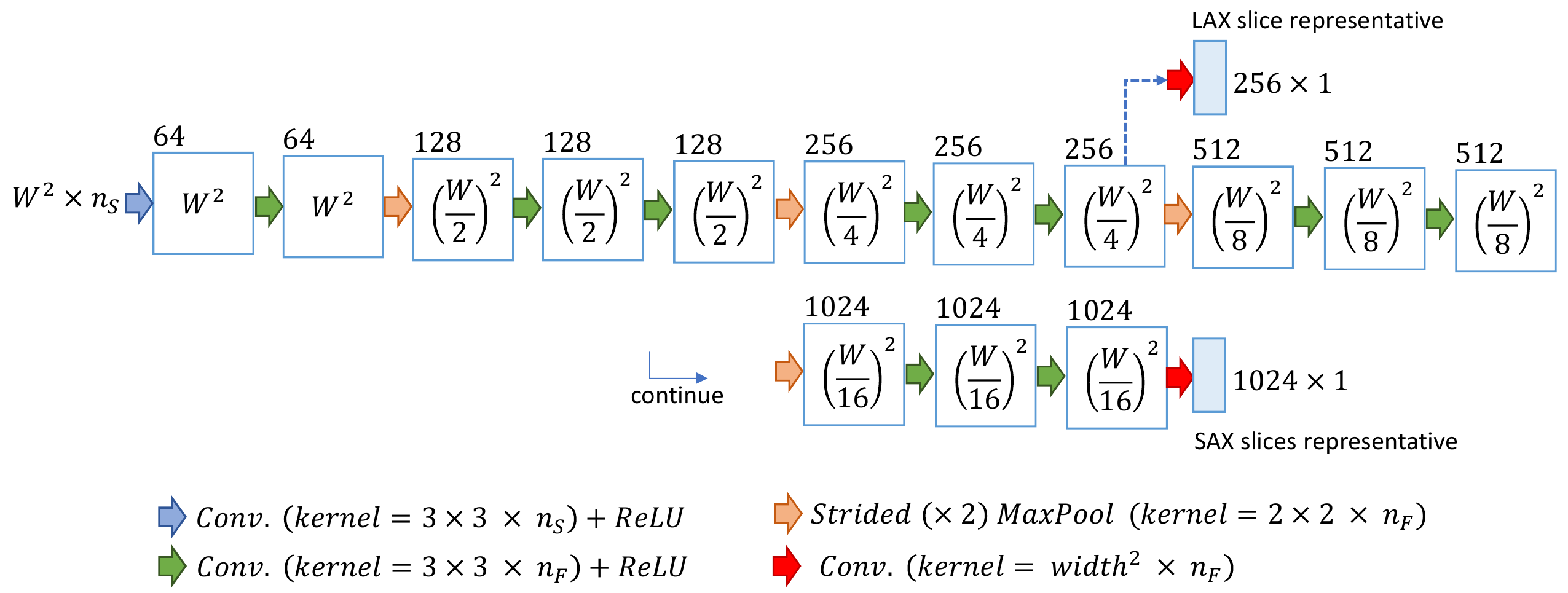}
	\vspace{-.3cm}
	\caption{The architecture of the two FCNs used in this study to obtain a vector of features representing the image information derived from the LAX and SAX views. A separate network is used for each view, with the LAX and SAX networks containing 9 and 15 layers respectively. The FCNs are composed of convolutional layers, with ReLUs and max-pooling. $W$ is the image width and height, $n_S$ and $n_F$ are the number of slices in each image volume and the number of activation maps respectively.}
		\label{CNN}
		\vspace{-0.5cm}
\end{figure*}

As illustrated in Fig.~\ref{CNN}, the FCN used in this work has been adapted from the down-sampling path of a U-Net~\cite{ronneberger2015u} architecture with an encoder depth of 2 for the LAX and 4 for the SAX images. The last layer of the FCNs have an extra convolutional layer with the kernel size of the current feature map dimensions to produce a vector of features - 1024 for SAX and 256 for LAX. The MLP has 11 inputs (size of metadata feature vector), 3 hidden layers (with 16, 32, and 64 neurons), and an output layer (with 128 neurons). ReLU is used in hidden and output layers. 

The outputs of the three sub-networks are concatenated to construct one feature vector (with the size of 1024+256+128=1408 neurons) that contains the behavioural, phenotypic, and demographic information derived from the metadata in addition to visual information from the imaging data. This information is fed into four fully connected layers, with ReLU (first two layers) and Sigmoid (last two layers) activation functions and batch normalisation, so that, by minimising $E(\theta)$ from Eq.~\ref{loss}, they produce the first $k$ parameters in PCA space which describe the 3D shape of the cardiac ventricles. To capture 99.7\% of shape variability in the training dataset we set $k=28$ and regress only those parameters from randomly initialised weights.

\section{Experiments and Results}
\subsection{Data and Annotations}
We performed experiments on 3500 CMR image volumes from the UK Biobank (UKB) using both end-diastolic and end-systolic time points. In terms of population sample size, experimental setup, and quality control, the most reliable reference annotations of cardiovascular structure and function found in the literature are those reported by~\cite{petersen2017reference}, in which CMR scans were manually delineated and analysed by a team of eight expert observers. These delineations were used to generate the reference 3D shapes, as explained in Sec.~\ref{Reference3DShapes}. The dataset was randomly split into a training (3000) and test set (500). The performance is reported on the test set with mean $\pm$ standard deviation.

\subsection{Implementation and Training}
The method was implemented using Python and Tensorflow. The network was trained using Adam for optimising the loss function (Eq.~\ref{loss}) with the learning rate of 0.001 and iteration number of 50,000 with a batch size of 10 subjects, all of which were determined empirically. There was no data augmentation. Training took $\sim$10 hours on Nvidia Tesla V100 GPUs hosted by Amazon Web Service and accessed using the MULTI-X platform~\cite{de2018multi}. At test time, it took less than a second to predict the shape parameters.

\subsection{Accuracy of Predicted Shapes}
\begin{figure*}[!t]
	\centering\includegraphics[width=0.7\linewidth]{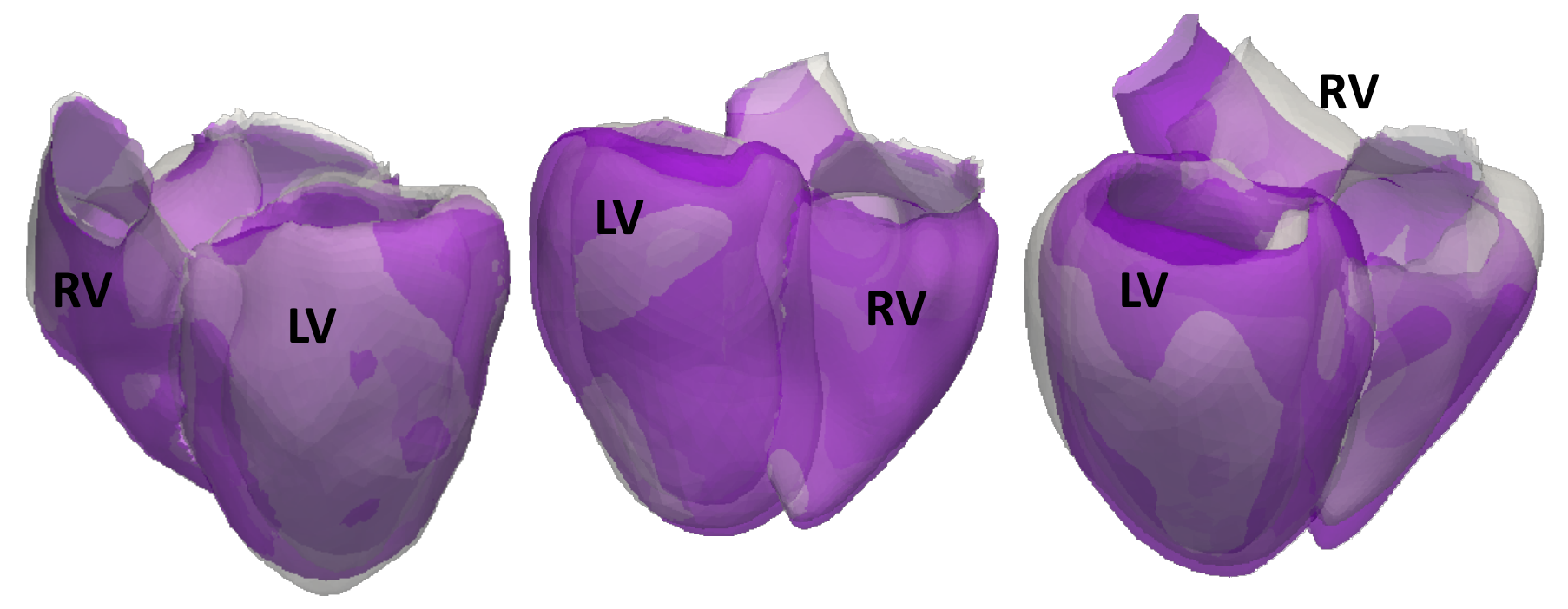}
	\vspace{-.4cm}
	\caption{Three samples of the generated 3D shapes of LV and RV. The gray shape is the reference whereas the purple shape is the predicted. }
		\label{samples}
		\vspace{-0.2cm}
\end{figure*}

Fig.~\ref{samples} shows some samples of ventricular shapes generated by our proposed method (in purple) overlaid with the corresponding reference shapes (in grey). It confirms that the network is capable of predicting accurate shape parameters to generate shapes very similar to the reference shapes obtained by manual delineations. To quantify the amount of similarity, we evaluated the performance of the proposed method by computing the Dice index ($\mathcal{D}$), and the mean ($\mathcal{M}$) and Hausdorff distance ($\mathcal{H}$) between reference and predicted shapes. Since this method outputs the parameters of a shape in the space, we first align the two shapes by removing their orientation and translation before computing the aforementioned metrics. $\mathcal{D}$ is between 0 and 1, with a higher $\mathcal{D}$ indicating a better match between the two shapes. $\mathcal{M}$ and $\mathcal{H}$ measure the mean and maximum distance, respectively, between the two surfaces, with a lower value indicating a better the agreement. Moreover, we report the effect of including the metadata in Table~\ref{IMG+MTDT}. As expected, the use of the metadata alongside the image information improves the network, leading to a more accurate prediction in all cardiac substructures. In addition to comparing against reference measurements, we also compare against one baseline method proposed by Attar et al.~\cite{attar} in which the authors carried out 3D analysis of the UKB CMR images using a shape model-based approach where the model is fitted during an iterative process using traditional intensity profiles.
\vspace{-0.5cm}

\begin{table*}[!ht]
\caption{Comparison of shape prediction accuracy using only images (IMG) or images with metadata (IMG+MTDT) in terms of $\mathcal{D}$(\%), $\mathcal{M}$(mm) and $\mathcal{H}$(mm) for LV endo-/epicardium and RV endocardium. \textbf{Bold} indicates best performing method.}
	\centering
	\scalebox{0.7}{
\begin{tabular}{cc|c|c|c|c|c|c|c|c} \cline{2-10} 
                        & \multicolumn{3}{c|}{\textbf{LV endocardium}} &  
                        \multicolumn{3}{c|}{\textbf{LV epicardium}} & 
                        \multicolumn{3}{c}{\textbf{RV endocardium}} \\ \cline{2-10} 
                        & \cite{attar} & IMG  & IMG+MTDT & \cite{attar} & IMG & IMG+MTDT & \cite{attar} & IMG & IMG+MTDT \\ \hline
\multicolumn{1}{c|}{\textbf{$\mathcal{D}$}} 
&\textbf{0.91$\pm$0.05} &0.82$\pm$0.09 &0.90$\pm$0.04 
&0.92$\pm$0.05  &0.83$\pm$0.08 &\textbf{0.93$\pm$0.05}
&0.88$\pm$0.08 &0.79$\pm$0.09 &\textbf{0.90$\pm$0.08}  \\ \hline

\multicolumn{1}{c|}{$\mathcal{M}$} 
&1.85$\pm$0.75  &3.45$\pm$0.94  &\textbf{1.81$\pm$0.70}  
&\textbf{1.80$\pm$0.62}  &3.02$\pm$0.84  &1.82$\pm$0.66 
&2.02$\pm$0.72  &3.00$\pm$0.91  &\textbf{2.00$\pm$0.70} \\ \hline

\multicolumn{1}{c|}{$\mathcal{H}$} 
&3.76$\pm$1.52  &8.78$\pm$1.96  &\textbf{3.11$\pm$1.49}  
&\textbf{3.32$\pm$1.38}  &7.69$\pm$1.77  &3.55$\pm$1.49 
&8.32$\pm$3.12  &12.11$\pm$5.21  &\textbf{7.05$\pm$3.03} \\ \hline
\label{IMG+MTDT}
\end{tabular}}
\vspace{-0.9cm}
\end{table*}

As shown in Table~\ref{IMG+MTDT}, $\mathcal{D}$ values show excellent agreement between reference and predicted shapes ($\geq 0.90$). $\mathcal{M}$ values are comparable to the in-plane pixel spacing range of 1.8 mm to 2.3 mm found in the UKB. Although $\mathcal{H}$ is larger, it is still within an acceptable range when compared with the distance range seen in~\cite{attar} or~\cite{duan2019automatic}. Note that the performance of the proposed method on RV is consistently better than the other approaches. Furthermore, we report the absolute and relative difference of the main cardiac function indexes (viz. LV and RV volume (ml) and myocardium mass (g)) derived from the predicted and the reference shapes in Table~\ref{clinical}. The proposed method achieved significantly lower error in volume and mass estimation, with p$<$0.001 in paired t-tests. 
\vspace{-0.6cm}

\begin{table*}[!]
\caption{Comparison of the absolute and relative difference between the reference and predicted shapes. \textbf{Bold} indicates best performing method.}
	\centering
	\scalebox{0.7}{
\begin{tabular}{cc|c|c|c|c|c} \cline{2-7} 
                        & \multicolumn{3}{c|}{\textbf{Absolute difference}} &  
                        \multicolumn{3}{c}{\textbf{Relative difference (\%)}} \\ \cline{2-7} 
                        & \cite{attar} & IMG  & IMG+MTDT & \cite{attar} & IMG & IMG+MTDT \\ \hline
\multicolumn{1}{l|}{LV Volume} 
&7.51$\pm$5.42  &9.80$\pm$6.33 &\textbf{6.01$\pm$4.98}  
&9.50$\pm$8.80  &10.31$\pm$9.45 &\textbf{8.03$\pm$5.05}  \\ \hline

\multicolumn{1}{l|}{LV Mass}
&8.42$\pm$5.22  &10.11$\pm$8.14  &\textbf{7.11$\pm$5.14}  
&9.10$\pm$8.01  &12.03$\pm$9.22  &\textbf{8.12$\pm$7.54} \\ \hline

\multicolumn{1}{l|}{RV Volume} 
&10.59$\pm$7.16  &12.62$\pm$10.14  &\textbf{9.24$\pm$5.20}  
&11.36$\pm$8.11  &14.55$\pm$9.89  &\textbf{10.03$\pm$7.00} \\ \hline

\label{clinical}
\end{tabular}}
\vspace{-0.7cm}
\end{table*}

Overall, the proposed method (IMG+MTDT) has superior accuracy to reference shapes than~\cite{attar}, while being on average $\sim$30 times faster during test time. This can be attributed to the combined use of image and patient metadata within a single network to directly predict shape parameters. The introduction of the metadata yielded a substantial positive impact on shape prediction with a $\sim$15\% average improvement across all metrics. We believe that including this information provides the network with a variable prior by allowing it to learn the likely distributions of shape parameters across different populations.

\section{Conclusion}
In this study, we presented a fully automatic framework capable of producing 3D cardiac shapes via the simultaneous use of images and patient metadata. We validated our workflow on a reference cohort of 500 subjects for which ground truth shapes exist with promising results. In particular, we showed a significant positive impact from including the metadata. As future work, in addition to investigating the effect of other clinical variables on shape prediction, we would like to increase the robustness of our pipeline to locate the shape in the image space, and handle severe pathological morphology, variable image quality, and alternative modalities. We also plan to explore the use of patient metadata in other deep learning applications.

\noindent
\newline
\textbf{{\large Acknowledgements}} \hspace{0.1cm}RA was funded by the School of Computing PhD Scholarship, University of Leeds. AFF acknowledges support from the Royal Academy of Engineering Chair in Emerging Technologies Scheme (CiET1819$\backslash$19) and the MedIAN Network (EP/N026993/1) funded by the Engineering and Physical Sciences Research Council (EPSRC).
\vspace{-0.15cm}
\bibliographystyle{ieeetr}
\bibliography{references.bib}
\end{document}